\documentstyle[amssymb]{elsart} \newcommand{\ket}[1]{|#1\rangle}
\begin{document}
\begin{frontmatter}
\title{On the general problem of quantum phase estimation} 
\author{G. M. D'Ariano, C. Macchiavello and M. F. Sacchi}
\address{Theoretical Quantum
Optics Group\\ Universit\`a degli Studi di Pavia and INFM Unit\`a di
Pavia\\ via A. Bassi 6, I-27100 Pavia, Italy}
\small{1998 PACS number(s): 03.65.-w, 03.65.Bz, 42.50.Dv, 42.50-p} 
\begin{abstract}
The problem of estimating a generic phase-shift experienced by a
quantum state is addressed for a generally degenerate phase shift operator.  
The optimal positive operator-valued measure is derived along with the 
optimal input state. Two relevant examples are analyzed: i) a multi-mode
phase shift operator for multipath interferometry; ii) the two mode
heterodyne phase detection.
\end{abstract}
\end{frontmatter}
\section{Introduction}
The problem of estimating the phase shift experienced by a radiation
beam has been the object of hundreds of studies in the last forty
years \cite{physcri}. The problem arises because for a single mode of
the electromagnetic field there is no selfadjoint operator for the
phase.  This is due to the semiboundedness of the number operator
\cite{shap-shep,ban} which is canonically conjugated to the phase as a
Fourier-transform pair \cite{shap93}. The most general and, at the same
time, concrete approach to the problem of the phase measurement is
quantum estimation theory \cite{helstrom}, a framework that has become
popular only in the last ten years in the field of quantum
information.  The most powerful method for deriving the optimal phase
measurement was given by Holevo \cite{Holevo} in the covariant case.
In this way the optimal positive operator-valued measure (POM) for
phase estimation has been derived for a single-mode field.  Regarding
the multi-mode case, only little theoretical effort has been spent
\cite{ban}, mostly devoting attention to the Lie algebraic structure
for two modes \cite{ban,ban2,luis}.  For two modes one can adopt the
difference between their photon numbers as the phase shift operator,
which thus is no longer bounded from below.  This opens the route
toward an exact phase measurement based on a selfadjoint operator
\cite{hradil}, with a concrete experimental setup using unconventional
heterodyne detection \cite{IEEE,max}.  The problem is however
complicated by the (infinite) degeneracy of the shift operator, and
for this reason the optimal states for this case have never been
derived.

In this paper the general problem of estimating the phase shift $\phi$
is addressed for any degenerate shift operator with discrete spectrum,
either $S=\mathbb Z$ (unbounded), or $S=\mathbb N$ (bounded from
below), or $S={\mathbb Z}_q$ (bounded), generalizing the Holevo method
for the covariant estimation problem. We find the optimal POM for
estimating the phase shift of a state $\ket{\psi_0}$, and then we
optimize the state itself. The degeneracy of the shifting operator is
removed through a simple projection technique.  The case of mixed
input state, which is generally very difficult, is considered in some
special situations. Two sections are devoted to the analysis of two
relevant examples: one concerning a multi-mode phase estimation problem
that arises in multi-path interferometry; the other involving a shift
operator that is the difference between the number of photons of two
modes, corresponding to unconventional heterodyne detection of the
phase.

\section{Optimal POM for the phase-shift estimation}
We address the problem of estimating the phase-shift $\phi $
pertaining to the unitary transformation
\begin{eqnarray}
\rho_\phi =e^{-i\phi \hat H}\,\rho_0\,e^{i\phi \hat H} \;\label{unit}
\end{eqnarray}
where $\hat H$ is a self-adjoint operator degenerate on the Hilbert space
${\cal H}$ with discrete
(un)bounded spectrum $S=\mathbb Z$, or $S=\mathbb N$, or $S={\mathbb Z}_q$,
$q>0$, and $\rho_0$ is a generic initial state (actually in the
following we will mostly restrict to the pure state case). 
The estimation problem is posed in the most general framework of quantum
estimation theory \cite{helstrom} on the basis of a cost function $C(\phi_*,
\phi)$ which weights the errors for the estimate $\phi_*$ given the true value
$\phi$. For a given {\it a priori} probability density $p_0(\phi)$ for the
true value $\phi$ the estimation problem consists in minimizing the average
cost
\begin{eqnarray}
\bar C=\int_0^{2\pi}d\phi \,p_0(\phi)\int_0^{2\pi}d\phi_*C(\phi_*,\phi)
\,p(\phi_*|\phi)\;,
\label{avc}
\end{eqnarray}
where $p(\phi_*|\phi)$ is the conditional probability of estimating
 $\phi_*$ given the true value $\phi$.  The average cost is minimized
 by optimizing the positive operator-valued measure (POM)
 \cite{helstrom} $d\mu(\phi_*)$ which gives the conditional
 probability by the Born rule
\begin{eqnarray}
p(\phi_*|\phi)d\phi_*=\mbox{Tr}[d\mu(\phi_*)e^{-i\phi \hat H}\rho_0 
e^{i\phi \hat H}]\;.
\label{prob}
\end{eqnarray}
We consider the general situation in which $\phi$ is {\it a priori}
uniformly distributed, i.e. with probability density
$p_0(\phi)=1/2\pi$.  Moreover, we want to weight errors independently
on the value $\phi$ of the phase, but only versus the size of the
error $\phi_*-\phi$, so that the cost function becomes an even
function of only one variable, i.e.  $C(\phi_*,\phi)\equiv
C(\phi_*-\phi)$.  It follows that also the optimal conditional
probability will depend only on $\phi_*-\phi$, and the optimal POM can
be obtained restricting attention only to phase-covariant POMs,
i.e. of the form
\begin{eqnarray}
d\mu(\phi_*)=e^{-i\hat H \phi_*}\xi e^{i\hat H\phi_*}\frac{d\phi_*}{2\pi}\;,
\label{dmu}
\end{eqnarray}
where $\xi$ is a positive operator.
satisfying the completeness constraints
needed for the normalization of
the POM $\int_0^{2\pi}d\mu(\phi)=1$.  
In fact, using Eq. (\ref{prob}) and the invariance of trace under
cyclic permutations one can easily recognize that
$p(\phi_*|\phi)\equiv p(\phi_*-\phi)$ if and only if $d\mu(\phi_*)$ is
covariant.  Hence the optimization problem resorts to finding the best
positive operator $\xi$ for a given cost function $C(\phi)$ and a
generic given state $\rho_0$.  As we will see, the POM obtained in
this way is optimal for a whole class of cost functions and initial
states $\rho_0$. Once the best POM is obtained, one further optimizes
the state $\rho_0$. This resorts to solving a linear eigenvalue
problem. In fact, the average cost can be written as the expectation
value of the cost operator $\hat C$, i.e.
\begin{eqnarray}
\bar C={\mbox{Tr}}[\hat C\rho_0]
\label{co}
\end{eqnarray}
where 
\begin{eqnarray}
\hat C=\int d\mu(\phi)C(\phi)\;.
\label{co2}
\end{eqnarray}
Using the Lagrange multipliers method to account for normalization and
mean energy one has to minimize the function
\begin{eqnarray}
{\cal L}[\rho_0]={\mbox{Tr}}[\hat C\rho_0]-\lambda{\mbox{Tr}}[\rho_0]
\label{elle}
\end{eqnarray}
which for a pure state $\ket{\psi_0}\langle\psi_0|$ is a quadratic form whose 
minimum is given by the eigenvalue equation
\begin{eqnarray}
\hat C\ket{\psi_0}=\lambda|\psi_0\rangle
\label{le} 
\end{eqnarray}
with the Lagrange parameter $\lambda$ playing the role of an
eigenvalue.  The linear problem can be easily extended to account also
for finite mean energy.

In summary, our problem is to minimize the cost $\bar C$ for a given
cost function $C(\phi)$ in Eq. (\ref{avc}). This is done in two steps:
i) by optimizing the positive operator $\xi$ for given generic fixed
state $\rho_0$: this will give a POM which is optimal for an
equivalence class of states ${\cal E}(\rho_0)$; ii) by further
optimizing the state in the equivalence class ${\cal
E}(\rho_0)$. Since the original state was arbitrarily chosen, this
will give the absolute minimum cost and the corresponding set of
optimal states and POM's.

The solution of the optimization problem is conveniently posed in the
representation where $\hat H$ is diagonal. The operator $\hat H$ is
generally degenerate, and we will denote by $\ket{n}_\nu$ a choice of
(normalized) eigenvectors corresponding to eigenvalue $n$, $\nu$ being
a degeneracy index, and by $\Pi_n$ the projector onto the
corresponding degenerate eigenspace. The problem for an input
generally mixed state $\rho_0$ is too difficult to address: therefore,
we focus our attention on the case of pure state
$\rho_0=\ket{\psi_0}\langle\psi_0|$, and we will leave some general
assertions on the mixed state case for the following. The problem is
restricted to the Hilbert space ${\cal H}_\parallel$ spanned by the
(normalized) vectors $\ket{n}\propto\Pi_n\ket{\psi_0}\neq 0$ with the
choice of the arbitrary phases such that $\langle n\ket{\psi_0}>0$.
Hence the POM can be chosen of the block diagonal form on ${\cal H}=
{\cal H_\parallel}\otimes{\cal H_\perp}$,
i.e. $d\mu(\phi)=d\mu_\parallel (\phi)\oplus d\mu_\perp(\phi)$ with
$d\mu_\perp(\phi)$ any arbitrary POM on ${\cal H} _\perp$. For the
optimization of the POM we consider $\Pi_n\ket{\psi_0}\neq 0$ $\forall
n\in S$, as it is clear that the resulting POM will be optimal also
for states having zero projection for some $n\in S$.  In this fashion
the problem is reduced to the ``canonical'' phase estimation problem
restricted to ${\cal H}_\parallel$: $\ket{\psi_0}\to \exp(i H_
\parallel\phi) \ket{\psi_0}$ where $H_\parallel =\sum_{n\in
S}n\ket{n}\langle n|$ and $\ket{\psi_0} =\sum_{n\in S}w_n\ket{n}$. Now
the problem is to find the positive operator $\xi_\parallel$ that
minimizes the cost $\bar C$ in Eq. (\ref{avc}).  On the $\ket{n}$
basis the operator $\xi_\parallel$ is written as
\begin{eqnarray}
\xi_\parallel=\sum_{n,m\in S}\ket{n}\langle m|\xi_{nm}\;.
\end{eqnarray}
For a generic even $2\pi$-periodic function
$C(\phi)=-\sum_{l=0}^{\infty}c_l\cos l\phi$ 
the average cost is given by
\begin{eqnarray}
\bar C=-c_0-\frac{1}{2}\sum_{l=1}^\infty c_l \sum_{|n-m|=l}
\langle{\psi_0}\ket{n}\langle m\ket{\psi_0}\xi_{nm}\;.
\label{avcxi2}
\end{eqnarray}
Positivity of $\xi$ implies the generalized Schwartz inequalities
\begin{eqnarray}
|\xi_{nm}|\leq\sqrt{\xi_{nn}\xi_{mm}}=1\;,
\label{mat}
\end{eqnarray}
where the last equality comes from the POM completeness $\int
d\mu_\parallel (\phi)= 1_{{\parallel}}$. One can write
\begin{eqnarray}
{\mbox{sign}}(c_l)
\sum_{|n-m|=l}\langle{\psi_0}\ket{n}\langle m\ket{\psi_0}\xi_{nm}
\leq
\sum_{|n-m|=l}|\langle{\psi_0}\ket{n}||\langle m\ket{\psi_0}|\;,
\label{inpom}
\end{eqnarray}
and the equality is obtained only for
$\xi_{nm}={\mbox{sign}}(c_{|n-m|})$ (notice that we chose
$\langle\psi_0\ket{n}> 0$ $\forall n\in S$).  The minimum cost is
\begin{eqnarray}
\bar C=-c_0 -\frac{1}{2}\sum_{l=1}^{\infty} |c_l|
\sum_{|n-m|=l}|\langle{\psi_0}\ket{n}||\langle m\ket{\psi_0}|
\label{minc}
\end{eqnarray}
where we put ${\mbox{sign}}(0)=1$, since the cost $\bar C$ is
independent of $\xi_{nm}$ for $c_{|n-m|}=0$.  Notice that positivity
of $\xi_\parallel$ is not generally guaranteed for any set of
${\mbox{sign}}(c_l)$. However, one can easily check that
$\xi_\parallel>0$ if
${\mbox{sign}}(c_{|n-m|})=\exp[i\pi(\epsilon_n-\epsilon_m)]$,
$\epsilon_n$ being any integer valued function of $n$. In fact, this
choice corresponds to a unitary transformation of the operator
$\xi_\parallel$ optimized with all $c_l\geq 0$ $\forall l\geq 1$ (the
parameter $c_0$ is irrelevant). The particular choice $c_l\geq 0$
$\forall l\geq 1$ has been considered by Holevo \cite{Holevo}, and
includes a large class of cost functions corresponding to the most
popular optimization criteria, as: {\em i)} the likelihood criterion
for $C(\phi)=-\delta_{2\pi}(\phi)$; {\em ii)} the $2\pi$-periodic
``variance'' for $C(\phi)=4\sin^2 (\phi/2)$; {\em iii)} the fidelity
optimization $C(\phi)=1-|\langle{\psi_0}|e^{i\hat
H\phi}\ket{\psi_0}|^2$ (here $c_l=2\sum_{|n-m|=l}|w_n|^2 |w_m|^2$).
For the Holevo class of cost functions the optimal POM becomes
\begin{eqnarray}
d\mu_\parallel(\phi)&=&\frac{d\phi}{2\pi} |e(\phi)\rangle \langle
e(\phi)|\label{pm2}\;,\label{pomopt}
\end{eqnarray}
where the (Dirac) normalizable vectors $|e(\phi)\rangle $
are given by
\begin{eqnarray}
|e(\phi)\rangle =\sum_{n\in S} e^{in\phi} |n\rangle \;.
\label{sg}
\end{eqnarray}
The vectors $|e(\phi)\rangle$ generalize the Susskind-Glogower
representation
$|e^{i\phi}\rangle=\sum_{n=0}^{\infty}e^{in\phi}|n\rangle$ for generic
integer spectrum. Therefore, the optimal POM $d\mu(\phi)$ is the
projector on the state $|e(\phi)\rangle$ in the Hilbert space
${\cal{H}}_\parallel$, and it is orthogonal for either $S=\mathbb Z$,
or $S={\mathbb Z}_q$, whereas it is not for $S=\mathbb N$.  Notice
that the POM (\ref{pomopt}) is also optimal for a density matrix
$\rho_0$ which is a mixture of states in ${\cal H}_\parallel$, with
the additional constraint of having constant phase along the
diagonals. This can be easily proved by re-phasing the basis $\ket{n}$
in such a way that all matrix elements of $\rho_0$ become
positive. Then the assertion easily follows in a way similar to the
derivation from Eq. (\ref{avcxi2}) to Eq. (\ref{minc}).  Moreover, it
is easy to see that the pure state case minimizes the cost, which for
the optimal POM is given by $\bar C=-\sum_{l=1}^\infty c_l \sum_{n\in
S}\langle n|\rho_0\ket{n+l}$ (remember that $\rho_0>0$ implies that
$|\langle n|\rho_0\ket{m}|^2\leq\langle n|\rho_0\ket{n} \langle
m|\rho_0\ket{m}$, and the bound is achieved by the pure state case
$\langle n|\rho_0\ket{m}=w_n^* w_m$).  Finally we want to emphasize
that for the bounded spectrum $S={\mathbb Z}_q$ there is no need for
considering a continuous phase $d\mu(\phi)$.  In fact, it is easy to
show \cite{oxf} that the same average cost is achieved by restricting
$\phi$ to the set of discrete values $\{\phi_s=\frac{2\pi
s}{q}\,,\quad s\in{\mathbb Z}_q\}$, ($q\equiv\mbox{dim}({\cal
H}_\parallel)$), and using as the optimal POM the orthogonal
projector-valued operator $|e(\phi_s)\rangle \langle e(\phi_s)|$.

Once the form of the optimal POM is fixed, one can optimize the state
$|\psi_0 \rangle $ solving the linear problem in Eq. (\ref{le}).  In
the following we show two examples of estimation of the phase shift
pertaining to highly degenerate integer operators (finite dimensional
cases are considered in Ref. \cite{oxf}).  In the first example we
consider the operator $\hat H=\sum_{l=1}^M l\,a_l^{\dag}a_l$ that
describes a multipath interferometer, involving $M$ different modes of
radiation. In the second, we focus our attention on the two-mode phase
estimation using unconventional heterodyne detection, where the phase
shift operator $\hat H=a^{\dag }a-b^{\dag }b$ is given by the
difference of photon numbers of the two modes.

\section{Optimal POM for multipath interferometer}
We consider the operator
\begin{eqnarray}
\hat H=\sum_{l=1}^M l\,a_l^{\dag}a_l\;\label{multi}
\end{eqnarray}
as the generator of the phase shift in Eq. (\ref{unit}). Such phase
shift affects a $M-$mode state of radiation in a multipath
interferometer, where contiguous paths suffer a fixed relative phase
shift $\phi $ \cite{multi} (this is also a schematic representation of
the phase shift accumulated by successive reflections in a Fabry-Perot
cavity). The operator $\hat H$ in Eq. (\ref{unit}) has integer
degenerate spectrum $S=\mathbb N$.  We can take into account the
degeneracy by renaming the number of photons of different modes as
follows
\begin{eqnarray}
\hat H|n\rangle_{\nu}= n|n\rangle_{\nu}\;,
\label{kappanu}
\end{eqnarray}
with $\nu=(\nu_2,\nu_3,\ldots,\nu_M)$, and
\begin{eqnarray}
|n\rangle_{\nu}\doteq\left|n-\sum_{l=2}^{M}l\nu_l\right\rangle\otimes
|\nu_2\rangle\otimes|\nu_3\rangle\otimes\ldots\otimes|\nu_M\rangle
\;.\label{bij}
\end{eqnarray}
The allowed values of $\nu $ are restricted to the set ${\cal E}_k$
given by
\begin{eqnarray}
{\cal E}_k&\doteq &\left\{\nu_2=0,1,\ldots,\left[{k\over2}\right],
\nu_3=0,1,\ldots, 
\left[\frac{k-2\nu_2}{3}\right]\;,\right.\nonumber \\& &\quad\ldots
\ ,\left.\ \nu_M=\left[\frac{k-\sum_{l=2}^{M-1}l\nu_l}{M}\right]\right\},
\end{eqnarray}
where $[x]$ denotes the integer part of $x$.  \par For the unshifted
initial state $|\psi_0 \rangle $ we choose a linear symmetrized
superposition of eigenvectors in Eq. (\ref{kappanu}), namely
\begin{eqnarray}
|\psi_0 \rangle =\sum_{n=0}^{\infty}w_{n}|n \rangle
_{\hbox{\scriptsize{sym}}}\;,
\label{fix}
\end{eqnarray}
where
\begin{eqnarray}
|n \rangle _{\hbox{\scriptsize{sym}}}&=&
\frac{1}{\sqrt{N_n}}\sum_{\{\nu_l\}}
\delta\left(\sum_{l=1}^Ml\nu_l-n\right)
|\nu_1\rangle\otimes|\nu_2\rangle\otimes\ldots\otimes|\nu_M\rangle\;,
\;\label{sym}
\end{eqnarray}
$N_{n}$ being the number of elements $\nu \in {\cal E}_n$.  Without
loss of generality, the basis $\ket{n}_{\hbox{\scriptsize{sym}}}$ has
been chosen such that the coefficients $w_n$ in Eq. (\ref{fix}) are
real and positive.  According to Eqs. (\ref{pomopt}) and (\ref{sg})
the optimal POM readily writes as follows
\begin{eqnarray}
d\mu(\phi)=\frac{d\phi}{2\pi}
\sum_{n,m=0}^{\infty}e^{i(n-m)\phi}\,
|n \rangle
_{\hbox{\scriptsize{sym}}}\ {}_{\hbox{\scriptsize{sym}}}\langle m|
\;.\label{pms}
\end{eqnarray}
One can now choose a cost function and then minimize the average cost
for the POM (\ref{pms}) upon varying the coefficients $w_n$ of the
state (\ref{fix}).  By choosing the cost function
$C(\phi)=4\sin^2(\phi/2)$ and by imposing the normalization constraint
through the Lagrange multiplier $\lambda $, the eigenvalue equation
(\ref{le}) gives the recursion for the coefficients $w_n$ of the form
\begin{eqnarray}
w_n +w_{n+2}-2 \lambda w_{n+1}=0  \;.\label{rec1}
\end{eqnarray}
The solutions of Eq. (\ref{rec1}) can be found in terms of the
Chebyshev's polynomials, and the corresponding optimal state writes as
follows
\begin{eqnarray}
|\psi \rangle=\left( \frac 2\pi \right)^{1/2}\sum_{n=0}^{\infty}
\sin[(n+1)\theta ]|n \rangle _{\hbox{\scriptsize{sym}}}\;,\qquad
\theta=\arccos \lambda \;.\label{cheb}
\end{eqnarray}
The state in Eq. (\ref{cheb}) is Dirac-normalizable. It is formally
equivalent to the eigenstate of the cosine operator $\hat C$ of the
phase of a single mode \cite{carr}.  The Dirac normalizability comes
from the non existence of normalizable states that minimize the
uncertainty relation for cosine and sine operators
\begin{eqnarray}
\Delta \hat C\,\Delta \hat S \geq \frac 12 
\left|\langle[\hat C,\hat S]\rangle \right |=\frac 14
\left\langle |0\rangle \langle 0|\right\rangle \;,
\end{eqnarray}
as proved in Ref. \cite{jack}.

\section{Phase-difference of two-mode fields}

In the previous example $\hat H$ was bounded from below and $S\equiv
\mathbb N$, such that the degenerate case is reduced to the standard
Holevo's problem. For the difference operator $\hat H=a^{\dag
}a-b^{\dag }b$ one has $S\equiv \mathbb Z$, and the set of
eigenvectors $|d\rangle _{\nu }$ can be written in terms of the joint
eigenvector $|n \rangle |m \rangle $ for the number operators $a^{\dag
}a$ and $b^{\dag }b$ with eigenvalues $n$ and $m$ as follows
\begin{eqnarray}
&&|d \rangle _{\nu }=|d+\nu \rangle |\nu \rangle\;, \nonumber \\& &
d\in {\mathbb Z}\,;\quad \nu \in [\max (0,-d),+\infty ) \;.\label{eid}
\end{eqnarray}
We consider an initial state $|\psi_0 \rangle $  of the form 
\begin{eqnarray}
|\psi _0\rangle =h_0 |0 \rangle |0 \rangle + \sum
_{n=1}^{+\infty}\left(h_n |n \rangle |0 \rangle + h_{-n}|0\rangle |n
\rangle \right) \;,\label{form}
\end{eqnarray}
where the basis has been chosen to have $h_n\geq 0$, $\forall n$.  The
optimal POM writes in the form of Eq. (\ref{pm2}) in terms of the
vectors $|\lambda_n\rangle$, $n\in\mathbb Z$, where
\begin{eqnarray}
|\lambda_n\rangle=\cases{|n\rangle_0\equiv |n\rangle |0\rangle\,,
\quad n\ge 0\;,\cr
|n\rangle_{|n|}\equiv |0\rangle ||n|\rangle\,, \quad n\le 0\;.\cr}
\end{eqnarray}
Here, the generalized Susskind-Glogower vector $|e(\phi)\rangle$ is
given by
\begin{eqnarray}
|e(\phi)\rangle =\sum_{n\in\mathbb Z}e^{in\phi}|\lambda_n\rangle\equiv
 |0 \rangle |0\rangle
+\sum _{d=1}^{+\infty}
\left(e^{id\phi}\,|d\rangle |0\rangle + e^{-id\phi
}|0 \rangle |d\rangle\right) \;.\label{sgd}
\end{eqnarray}
Notice that, differently from the usual case of spectrum $S=\mathbb
N$, now the POM is orthogonal (in the Dirac sense):
\begin{eqnarray}
\langle e(\phi)|e(\phi ')\rangle=\sum_{n=-\infty}^{+\infty}
e^{in(\phi-\phi ')}=\delta_{2\pi}(\phi-\phi ')\;,
\end{eqnarray}
where $\delta_{2\pi}(\phi)$ is the Dirac comb. This means that in this
case it is possible to define a selfadjoint phase operator
\begin{eqnarray}
\hat\phi=\int_{-\pi}^{+\pi}d\phi |e(\phi)\rangle\langle e(\phi)| \phi\;,
\end{eqnarray}
as already noticed by Hradil and Shapiro \cite{hradil,IEEE}.

We now address the problem of finding the normalized state of the form
(\ref{form}) with finite mean photon number that minimizes the average
cost evaluated through the ideal POM (\ref{pm2}).  As a cost function
we choose again $C(\phi)=4\sin^2(\phi/2)$ (periodicized-variance
criterion), corresponding to the cost operator
\begin{eqnarray}
\hat C=2-e^+-e^-\;,\label{hatc}
\end{eqnarray}
where
\begin{eqnarray}
e^+=\sum _{n\in\mathbb Z}|\lambda_{n+1}\rangle \langle\lambda_n|\;,
\qquad e^- =(e^+)^{\dag}\;.
\end{eqnarray}
Introducing the energy operator $\hat E=a^\dag a+b^\dag b$ and an
additional Lagrange parameter accounting for finite mean energy
$\langle\hat E\rangle$, the eigenvalue problem in Eq. (\ref{le})
rewrites as follows
\begin{eqnarray}
[\hat C-\lambda '-\mu '(a^\dag a+b^\dag b)]\ket{\psi_0}=0\;,
\end{eqnarray}
where $\lambda '$ and $\mu '$ are the Lagrange multipliers for
normalization and mean energy, respectively.  The following recursion
relations for the coefficients $h_n$ is obtained
\begin{eqnarray}
h_{n+1}+h_{n-1}-\mu (\lambda +|n|)h_n=0\;,\label{rec}
\end{eqnarray}
with $\lambda =(\lambda '-2)/\mu '$ and $\mu =-\mu '$.  The solution
of Eq. (\ref{rec}) is given in terms of Bessel functions of the first
kind in the following form
\begin{eqnarray}
h_n=k(\lambda ,\mu )\,J_{\lambda +|n|}(2/\mu)
\;,\label{hn}
\end{eqnarray}
$k(\lambda ,\mu)$ being the constant of normalization
\begin{eqnarray}
k(\lambda ,\mu)=\left[\sum _{n=-\infty} ^{+\infty} J^2_{\lambda
+|n|}(2/\mu)\right]^{-1/2} \;.\label{norm}
\end{eqnarray}
The matching of the recursion for positive and negative indices leads
to the condition
\begin{eqnarray}
\lambda J_{\lambda }(2/\mu ) -(2/\mu)J_{\lambda +1}(2/\mu )=
(2/\mu)\frac{\hbox{d}}{\hbox {d}\,(2/\mu )}J_{\lambda }(2/\mu )=0\;.
\label{lmu}\end{eqnarray} 
Eq. (\ref{lmu}) has infinitely many solutions $\mu =\mu(\lambda )$,
and one needs to further minimize the average cost in Eq. (\ref{avc})
versus the average photon number $N$ parameterized by $\lambda $ and
$\mu=\mu (\lambda )$
\begin{eqnarray}
N=2k(\lambda ,\mu)^2 \left[\sum _{n=0} ^{+\infty} n\,J^2_{\lambda
+n}(2/\mu)\right] \;.
\end{eqnarray}
In this way one can find the normalized and finite-energy states that
achieve the minimum cost for the optimal POM.  \par The solution
(\ref{hn}) of the recursive relation (\ref{rec}) has some similarity
with the solution for the minimum phase-uncertainty states of a
single-mode field \cite{carr,jack}. The proof of convergence of the
series in Eq. (\ref{norm}) can be found in Ref. \cite{jack}. However,
the matching condition (\ref{lmu}) (instead of the vanishing condition
for $h_n$ with $n<0$ for one mode) makes the two-mode phase estimation
problem more difficult, since one cannot exploit the properties of the
zeros of the Bessel functions in an asymptotic approximation, as done
in Ref. \cite{band} for the single-mode case.

 
\end{document}